\documentclass[aps,amsmath,amssymb,prc,twosides,twocolumn,floats,superscriptaddress]{revtex4-2}

\usepackage{graphicx} 
\usepackage[utf8]{inputenc}
\usepackage[english]{babel}
\usepackage[shortlabels]{enumitem}
\usepackage{bm}
\usepackage{float}
\usepackage{tabularx}
\usepackage{array}
\usepackage[dvipsnames]{xcolor}
\usepackage{hyperref}

\usepackage{multirow}
\usepackage{longtable}
\usepackage{dcolumn}
\newcolumntype{d}{D{.}{.}{-1}}
\usepackage{setspace}
\usepackage{threeparttable}
\usepackage{tabularx}
\usepackage[symbol*]{footmisc}
\DefineFNsymbolsTM{myfnsymbols}{
  \textasteriskcentered *
  \textdagger    \dagger
  \textdaggerdbl \ddagger
  \textsection   \mathsection
  \textbardbl    \|%
  \textparagraph \mathparagraph
}%
\setfnsymbol{myfnsymbols}

\usepackage{siunitx}
\newcommand{\nuc}[2]{\hbox{$^{#1}$#2}}

\usepackage{gensymb}
\usepackage{mathtools}

\begin{document}

\title{Excited states of \nuc{148}{Nd} studied via the $\nuc{150}{Nd}(p,t)\nuc{148}{Nd}$ reaction and the observation of possible low-spin two-phonon octupole states at $N=88$}

\author{A.L.~Conley}
\affiliation{Department of Physics, Florida State University, Tallahassee, Florida 32306, USA}

\author{M.~Spieker}
\email{Corresponding author: mspieker@fsu.edu}
\affiliation{Department of Physics, Florida State University, Tallahassee, Florida 32306, USA}

\author{R.~Aggarwal}
\affiliation{Department of Physics, Florida State University, Tallahassee, Florida 32306, USA}

\author{L.T.~Baby}
\affiliation{Department of Physics, Florida State University, Tallahassee, Florida 32306, USA}

\author{J.~Davis}
\affiliation{Department of Physics, Florida State University, Tallahassee, Florida 32306, USA}

\author{J.~Esparza}
\affiliation{Department of Physics, Florida State University, Tallahassee, Florida 32306, USA}

\author{I.~Hay}
\affiliation{Department of Physics, Florida State University, Tallahassee, Florida 32306, USA}

\author{B.~Kelly}
\affiliation{Department of Physics, Florida State University, Tallahassee, Florida 32306, USA}

\author{T.~Kirk}
\affiliation{Department of Physics, Florida State University, Tallahassee, Florida 32306, USA}

\author{M.I.~Khawaja}
\affiliation{Department of Physics, Florida State University, Tallahassee, Florida 32306, USA}

\author{R.~Mahajan}
\altaffiliation{Current address: Department of Physics and Astronomy, University of Kentucky, Lexington, Kentucky 40506-0055, USA}
\affiliation{Department of Physics and Astronomy, Louisiana State University, Baton Rouge, Louisiana 70803, USA}

\author{M.~Mestayer}
\affiliation{Department of Physics, Florida State University, Tallahassee, Florida 32306, USA}

\author{A.B.~Morelock}
\affiliation{Department of Physics and Astronomy, University of Tennessee, Knoxville, Tennessee 37996, USA}

\author{A.~Peters}
\affiliation{Department of Physics, Florida State University, Tallahassee, Florida 32306, USA}

\author{A.M.~Ring}
\affiliation{Department of Physics, Florida State University, Tallahassee, Florida 32306, USA}

\author{J.~Sheridan}
\affiliation{Department of Physics, Florida State University, Tallahassee, Florida 32306, USA}

\author{V.~Sitaraman}
\affiliation{Department of Physics, Florida State University, Tallahassee, Florida 32306, USA}

\author{T.~Stuck}
\affiliation{Department of Physics, Florida State University, Tallahassee, Florida 32306, USA}

\date{\today}

\begin{abstract}

We report new data from a $\nuc{150}{Nd}(p,t)\nuc{148}{Nd}$ experiment performed at the John D. Fox Accelerator Laboratory of Florida State University. In total, 54 excited states of \nuc{148}{Nd} were observed up to an excitation energy of 3500\,keV. In this work, we focus on $0^+$ states and their band members. In contrast to previous work, the $0^+_3$ band is proposed as the candidate for the two-phonon octupole vibrational band. Supporting $spdf$ IBM-1 calculations are presented. To test the robustness of the IBM calculations, several observables were interrogated and are discussed in this publication. In addition, we make the case that neither the $0^+_2$ nor the $0^+_3$ states of the other $N=88$ isotones are likely good candidates for two-phonon octupole states. Based on our new data for \nuc{148}{Nd}, we propose candidates in \nuc{150}{Sm} and \nuc{152}{Gd}. Using available $\gamma$-decay data for states with moderate spins in the yrast sequence and a comparison to IBM calculations, we also show how the staggering of the $B(E1)/B(E2)$ ratios in the yrast sequence can possibly be used to probe the appearance of bands with multiple octupole phonons.

\end{abstract}


\maketitle

\section{Introduction}

Nd $(Z=60)$, Sm $(Z=62)$, and Gd $(Z=64)$ isotopes around neutron numbers $N=88$ and $N=90$ have attracted the interest of the nuclear-structure community for decades and their detailed study considerably shaped our understanding of the emergence of collectivity in atomic nuclei in general. Some examples include the emergence of quadrupole deformation beyond $N=90$ discussed within the scope of critical-point symmetries \cite{Iac01a, Kru02a, Cas06a, Cas09a, Cej10a}, enhanced octupole collectivity at $N=88$ \cite{Ibb93a, But96a, Ibb97a, Buc16a, Buc17a, Nom21a, Pas25a}, contributions of triaxial degrees of freedom \cite{Ots25a} recently studied via, {\it e.g.}, the $\gamma$ decay of the isovector giant dipole resonance in \nuc{154}{Sm} \cite{Kle25a}, the possible appearance of $\alpha$ clustering in lanthanides studied via enhanced $E1$ transitions \cite{Spi15a} and $\alpha$-transfer reactions \cite{Jae82a}, and the study of pairing properties with special focus on $J^{\pi}=0^+$ states and their connection to collective degrees of freedom via two-neutron transfer reactions \cite{Mey06a, Cla09a, Zha17a, Nom19a, Apr25a, Nom26a}.

\begin{figure*}
    \centering
    \includegraphics[width=0.99\linewidth]{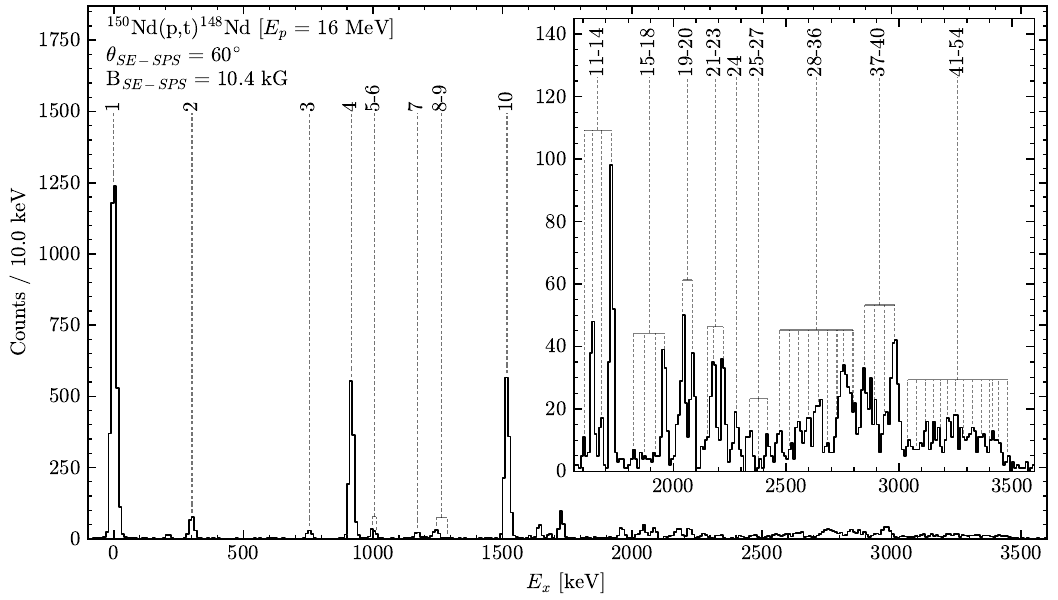}
    \caption{Excitation-energy spectrum of \nuc{148}{Nd} measured in the $\nuc{150}{Nd}(p,t)\nuc{148}{Nd}$ reaction with the SE-SPS at a laboratory scattering angle of $\theta_{SE-SPS}=60^{\circ}$ and with the magnetic field set to 10.4\,kG for the spectrograph. States identified as excited states of \nuc{148}{Nd} are marked with numbers. Note the different scale of the inset. See the table in the supplemental material \cite{suppl} for a list of states observed in the $(p,t)$ reaction.}
    \label{fig:spectrum}
\end{figure*}

In this publication, we present new data obtained from a $\nuc{150}{Nd}(p,t)\nuc{148}{Nd}$ experiment performed at the John D. Fox Accelerator Laboratory of Florida State University with the Super-Enge Split-Pole Spectrograph \cite{spi24a}. Several excited states of \nuc{148}{Nd} $(N=88)$ were identified, their $(p,t)$ angular distributions measured, and $(p,t)$ angle-integrated cross sections determined. Limited data from previous $(p,t)$ experiments exist \cite{Max66a, Yag72a}. These data suggested the existence of $0^+$ states with significant $(p,t)$ cross sections beyond the $0^+_2$ state. However, partly because of remaining ambiguities and data unavailability, data for \nuc{148}{Nd} were not included in the systematic comparison of relative $(p,t)$ intensities presented in Refs.\,\cite{Zha17a, Nom19a, Nom26a}. These theoretical studies connected the evolution of $(p,t)$ and $(t,p)$ intensities along isotopic chains to the phase transition from spherical to axially deformed shapes at $N=90$. Here, we add the missing data for \nuc{148}{Nd}. In addition, we address the open question whether the $0^+_2$ states of the $N=88$ isotones are indeed of two-phonon octupole structure as predicted in Refs.\,\cite{Nom21a, Nom26a}. Earlier, it had already been suggested that the large number of observed $0^+$ states in the deformed lanthanides could possibly be explained by adding two-phonon octupole excitations to the model space \cite{Zam02a}. Similar studies in the actinides followed, see, {\it e.g.}, Refs.\,\cite{Zam03a, Wir04a, Lev13a, Spi13a, Spi18a} and references therein. For the specific case of \nuc{148}{Nd}, Ibbotson {\it et al.} had observed enhanced $E3$ transitions linking the negative-parity band to states of the band built on the $0^+_2$ state \cite{Ibb93a, Ibb97a}, which the authors called the $\beta$ band at the time. They stated that this observation might be indicative of significant admixtures of the two-phonon octupole vibration to the wavefunctions of the $\beta$ band members \cite{Ibb93a, Ibb97a}. The work of Refs.\,\cite{Bvu13a, Zim16a} extended this idea to the $N=88$ isotones \nuc{150}{Sm}, \nuc{152}{Gd}, and \nuc{154}{Dy} by observing enhanced $E1$ transitions between members of the $0^+_2$ band and the negative-parity band. It has to be noted though that the $0^+_2$ band head is at lower energies than the corresponding band head of the negative-parity band, which might challenge a simple two-phonon octupole interpretation of the positive-parity band. The authors of Refs.\,\cite{Bvu13a, Zim16a} mentioned a shape-coexistence scenario similar to the one proposed for actinide nuclei in Ref.\,\cite{Cha79a}. In that case, the $0^+_2$ band would belong to a configuration that is octupole deformed and different from the ground-state configuration. We also note that the case of previously discussed two-phonon octupole states in the $N=88$ isotones is different from the high-spin two-phonon octupole vibrational states reported for, {\it e.g.}, \nuc{148}{Gd} \cite{Lun84a, Pod00a}, \nuc{144}{Nd} and \nuc{146}{Sm} \cite{Bar95a}.

\section{Experimental Details}

The $\nuc{150}{Nd}(p,t)\nuc{148}{Nd}$ experiment was performed using the Super-Enge Split-Pole Spectrograph (SE-SPS) and its light-ion, position-sensitive focal-plane detector at the John D. Fox Accelerator Laboratory of Florida State University \cite{spi24a}. Negatively charged hydrogen ions were injected from a SNICS source into the laboratory's Super-FN Tandem Van de Graaff accelerator, accelerated to 16 MeV, and delivered to the SE-SPS, where they impinged onto a 150-$\mu$g/cm$^2$ thick and 98\,$\%$ enriched \nuc{150}{Nd}  target layer evaporated onto a 25-$\mu$g/cm$^2$ thick Carbon backing. Differential cross sections, $d\sigma/d\Omega$, were measured at nine laboratory scattering angles ranging from 10$^{\circ}$ to 60$^{\circ}$. Depending on the scattering angle, the SE-SPS magnetic field was either set to 9.7\,kG or 10.4\,kG to cover excitation energies from 0\,keV to about 3500\,keV for \nuc{148}{Nd}. Well-known, low-lying excited states with energies up to 2000\,keV were used for the energy calibration of the triton spectra. An example spectrum is shown in Fig.\,\ref{fig:spectrum}, highlighting states of \nuc{148}{Nd} which were populated in the $(p,t)$ reaction. Several states could be identified. Above 3\,MeV of excitation energy, the density of states complicated the clear identification of individual states. To identify these, the full width at half maximum (FWHM) was kept constant when fitting the spectrum and extracting yields for the differential cross sections. For this experiment, the FWHM was between 30 and 40\,keV, depending on the scattering angle and with a solid-angle coverage of $\Delta \Omega=4.6$\,msr. Corrections for higher-order aberrations, as briefly described in Ref.\,\cite{spi24a}, were not performed. A full table with all states observed in the $(p,t)$ reaction is provided with the supplemental material \cite{suppl}.

\begin{figure*}[t]
    \centering
    \includegraphics[width=0.99\linewidth]{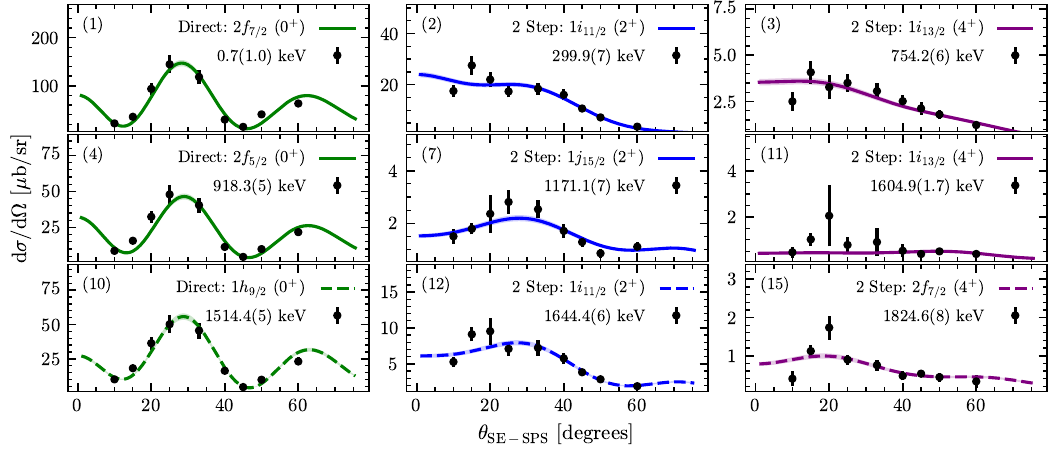}
    \caption{$(p,t)$ angular distributions measured for states in \nuc{148}{Nd} associated with the band built on the $0^+_1$ state (states \#1, \#2, and \#3 in Fig.\,\ref{fig:spectrum}), the states of the $0^+_2$ band (states \#4, \#7, and \#11), and the tentatively assigned band built on the $0^+_3$ state (states \#10, \#12, and \#15). Lines correspond to DWBA calculations performed with the coupled-channels program \textsc{chuck3}\,\cite{chuck} The theoretical distributions were scaled to the data by minimizing $\chi^2$. The uncertainty bands indicate the $1\sigma$ confidence interval. For excited states without an adopted spin-parity assignment, or for states that have not been previously reported, the associated theoretical distributions are shown with dashed lines.}
    \label{fig:first_bands}
\end{figure*}

\section{Results and Discussion}

A total of 54 excited states were observed. For all of these, angular distributions were measured. These distributions and the angle-integrated cross sections are presented in the supplemental material \cite{suppl}. However, only for 18 of the states populated in $(p,t)$, we attempted to fit the measured distributions using coupled-channels calculations. To do so, theoretical $(p,t)$ angular distributions were calculated with the coupled-channels program \textsc{chuck3}\,\cite{chuck} using the distorted wave Born approximation (DWBA), and the global optical-model parameters of Ref.\,\cite{Kon03a} for protons and of Ref.\,\cite{Bec69a} for tritons. For states with $J \neq 0$, two-step processes had to be included. In this work, only the simplest two-step scheme was considered, in which the reaction proceeds through the $2^+_1$ state in \nuc{150}{Nd} at 130\,keV. For deformed nuclei with comparably low-lying excited (collective) states, such two-step processes had to be included in many previous $(p,t)$ studies to obtain agreement with measured angular distributions, see, {\it e.g.}, Refs.\,\cite{Bae73a, Lev09a, Spi18a, Buc23a}.

Example angular distributions for the $0^+$, $2^+$, and $4^+$ states belonging to the three lowest-lying bands built on $0^+$ states are shown in Fig.\,\ref{fig:first_bands}. For the $0^+$ states, the agreement between the measured and DWBA angular distributions assuming direct transfer of the two neutrons is exceptional. For the low-lying $2^+$ and $4^+$ states, multi-step contributions clearly change the angular distribution. Using the simple scheme where the reaction proceeds through the $2^+_1$ state of \nuc{150}{Nd}, reasonable agreement between experiment and theory is obtained. The neutron orbitals from which the neutrons get transferred in a $\Delta s = 0$ state are also indicated. Of the states shown in Fig.\,\ref{fig:first_bands}, the ground state and the excited $0^+$ state at 917\,keV, the $2^+$ states at 302\,keV and 1171\,keV, and the $4^+$ states at 752\,keV and 1604\,keV are adopted \cite{NDS148}. As can be seen in Fig.\,\ref{fig:first_bands}, the excitation energies determined from our data agree within 2\,keV with the adopted energies.

\begin{figure}[t]
    \centering
    \includegraphics[width=0.99\linewidth]{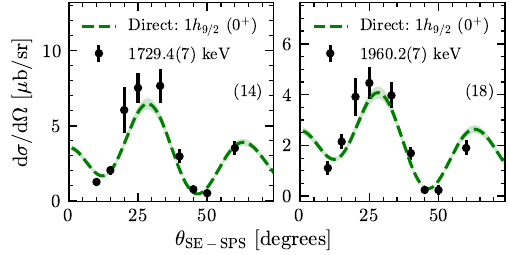}
    \caption{Same as Fig.\,\ref{fig:first_bands} but for the other $0^+$ states (states \#14 and \#18 in Fig.\,\ref{fig:spectrum}) observed in the $\nuc{150}{Nd}(p,t)\nuc{148}{Nd}$ reaction.}
    \label{fig:0+states}
\end{figure}

The characteristic $\ell =0$ $(p,t)$ angular distribution, observed for the $0^+_1$ and $0^+_2$ states in Fig.\,\ref{fig:first_bands}, allowed us to unambiguously identify three additional excited $0^+$ states at 1514.4(5)\,keV, 1729.4(7)\,keV, and 1960.2(7)\,keV. See Figs.\,\ref{fig:first_bands} and \ref{fig:0+states}. Earlier, Maxwell {\it et al.} had observed a strongly populated state at 1.58\,MeV in their $\nuc{150}{Nd}(p,t)\nuc{148}{Nd}$ study\,\cite{Max66a}, but did not report an angular distribution. Comparing their lower-resolution $(p,t)$ spectrum to ours, it is likely that this state is the state which we observe at 1514\,keV (state \#10 in Fig.\,\ref{fig:spectrum}). This 1514-keV $0^+$ state possibly corresponds to the state adopted at 1516\,keV \cite{NDS148}. It decays via a 492-keV $\gamma$-ray transition to the $1^-_1$ state, which was observed in the $\beta^-$ decay of \nuc{148}{Pr} \cite{Kar88a}. This $\gamma$ decay would then correspond to an $E1$ transition. We will get back to the significance of such an $E1$ decay to the one-phonon octupole vibrational band. The other two excited $0^+$ states likely do not coincide with any of the currently adopted states of \nuc{148}{Nd}.

The state at 1644.4(6)\,keV, for which we propose a $J^{\pi} = 2^+$ assignment, likely corresponds to the state currently adopted at 1646\,keV \cite{NDS148}. This state is known to decay to the $0^+_1$, $2^+_1$, and $1^-_1$ states, which further supports our proposed $J^{\pi} = 2^+$ assignment.

The state, which we observe at 1824.6(8)\,keV and for which we suggest a $J^{\pi} = 4^+$ assignment, may correspond to the state currently adopted at 1825\,keV \cite{NDS148}. For this state, $\gamma$-ray transitions to the $4^+_1$ and $3^-_1$ state are listed \cite{NDS148}, which would also agree with the suggested spin-parity assignment.

\begin{figure}[t]
    \centering
    \includegraphics[width=0.99\linewidth]{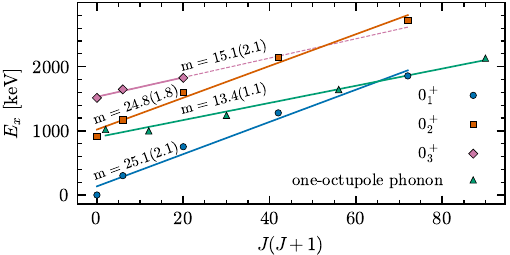}
    \caption{Low-lying bands recognized in \nuc{148}{Nd}. Bands built on the $0^+_1$ and $0^+_2$ states, as well as the one-phonon octupole vibrational band are already adopted \cite{NDS148}. The band built on the 1514-keV, $0^+_3$ state is proposed in this work. The slopes, $m$, determined from a linear fit to these sequences are given with the associated $1\sigma$ confidence interval. As will be discussed in the text, we propose to recognize the tentatively assigned $6^+$ and $8^+$ states at 2149\,keV and 2726\,keV \cite{Ibb97a}, respectively, as band members of the $0^+_3$ rather than the $0^+_2$ band.}
    \label{fig:rotation}
\end{figure}

For the $0^+_1$ and $0^+_2$ bands, their $2^+$ and $4^+$ members were already placed \cite{NDS148}. As indicated in Fig.\,\ref{fig:first_bands}, we propose to recognize the 1644-keV and 1825-keV states as members of a band built on the 1514-keV $0^+_3$ state. To support this assignment, we compare this new band to the bands mentioned above and also include the one-phonon octupole vibrational band in Fig.\,\ref{fig:rotation}. The suggested members of the $0^+_3$ band clearly follow the linear trend of increasing excitation energy expected for a rotational sequence when plotted against $J(J+1)$. The slope determined for this $0^+_3$ band is  significantly closer to the one of the one-phonon octupole vibrational band than to the one of the ground-state band. Interestingly, the slope determined for the $0^+_2$ band is much closer to the one of the ground-state band. The similarities of the associated moments of inertia might hint at structure similarities between the respective bands. Of course, to uniquely establish band assignments, intraband transitions would need to be detected.

As mentioned above, all low-spin states of the proposed $0^+_3$ band decay to the one-phonon octupole vibrational band. In combination with comparable moments of inertia, such $E1$ decays to the one-phonon octupole vibrational states were used in Refs.\,\cite{Lev13a, Spi18a} to argue in favor of the two-phonon octupole interpretation for low-lying $0^+$ rotational bands. The $B(E1)/B(E2)$ \cite{Spi18a} ratios in \nuc{148}{Nd} are also enhanced. The values, derived from the evaluated data \cite{NDS148}, have been compiled in Table\,\ref{tab:01}. Multipole mixing ratios are not known to quantify any $M1$ contributions where necessary. Note that these ratios are different from the ones discussed in Refs.\,\cite{Bvu13a, Zim16a} though, where the $E2$ transitions were intraband rather than interband transitions. As mentioned earlier, the $0^+_2$ is energetically lower than the $3^-_1$ and $1^-_1$ states in \nuc{148}{Nd}. No $\gamma$ decay of its $2^+$ band member to the negative-parity band has been observed so far \cite{NDS148}. For the 1604-keV $4^+$ state, a 605-keV $\gamma$-ray transition to the $3^-_1$ state is adopted, however, without any reported decay intensity \cite{NDS148}. A $B(E1)/B(E2)$ ratio can be calculated from the matrix elements reported in Ref.\,\cite{Ibb97a}, see Table\,\ref{tab:01}. Based on the generally enhanced ratios in the $0^+_3$ band, it seems more likely that this band would be the candidate for a two-phonon octupole band. We will provide more evidence. A critical missing piece is its $6^+$ band member. Earlier, Ibbotson {\it et al.} reported the observation of a tentative $6^+$ state at 2149\,keV and placed it in the band built on the $0^+_2$ state \cite{Ibb93a, Ibb97a}. We suggest to interpret it as a band member of the band built on the $0^+_3$ state instead. This proposal is based on the fact that this state also shows enhanced $E1$ decays to the one-phonon octupole vibrational band \cite{Ibb97a} (see Table\,\ref{tab:01}) and that it fits into the band sequence built on the $0^+_3$ state (see Fig.\,\ref{fig:rotation}). We will also show that the $E3$ matrix elements reported for that state can be explained in a simplistic model that supports such a band assignment. The $6^+$ belonging to the $0^+_2$ band could be the state adopted at 2099\,keV, which would also fit into its band sequence and for which no decays to the one-phonon octupole vibrational band have been observed so far \cite{NDS148}. That, of course, raises the question why the $4^+$ state at 1604 keV of that band decays to the one-phonon octupole vibrational band. A careful inspection of the $B(E1;4^+_{1604} \rightarrow 3^-_1)$ value shows that it is an order of magnitude smaller than the $B(E1;6^+_{2149} \rightarrow 5^-_1)$ value, see Table\,\ref{tab:01}. Still, the $B(E1)/B(E2)$ ratio is as large as for members in the $0^+_3$ band, i.e., as for the candidate for the two-phonon octupole band. This emphasizes that both the $B(E1)$ and $B(E2)$ transition rates are important for structure assignments. Some $B(E2)$ strengths were added to Table\,\ref{tab:01}. In the case of the 1604-keV $4^+$ state, we are in fact dealing with small $B(E2)$ transition strengths to the ground-state band in addition to a less enhanced $E1$ transition. For reference, the $B(E2;4^+_{1604} \rightarrow 2^+_1)$ is significantly smaller than 1 W.u. and the $B(E2;6^+_{2149} \rightarrow 4^+_1)$ is also only around 4 W.u. \cite{Ibb97a}, see Table\,\ref{tab:01}. Both of them are clearly non-collective transitions. The $B(E2;6^+_{2149} \rightarrow 6^+_1)$ value, calculated from the $E2$ matrix element reported in Ref.\,\cite{Ibb97a}, is, however, indicative of a collective transition.

\begingroup
\squeezetable
\renewcommand*{\arraystretch}{1.2}
\begin{table*}[!t]
\caption{\label{tab:01} Comparison of experimental data for the two-phonon octupole candidates and predictions of the $spdf$ IBM-1 using the parameters of Ref.\,\cite{Spi15a, Spi16a} for \nuc{148}{Nd}. As in Refs.\,\cite{Lev13a, Spi13a}, the dipole interaction $\hat{D}^{\dagger}_{spdf}\hat{D}_{spdf}$ was added to the Hamiltonian with a small coupling strength of $\alpha = 0.001$\,MeV. Without it, no $E2$ transitions would be possible between states with two negative-parity bosons and no negative-parity bosons in their wavefunctions. The admixture of $f$ and $p$ bosons to the ground state of \nuc{148}{Nd} is small, i.e., $\langle \hat{n}_f \rangle + \langle \hat{n}_p \rangle \approx 0.004$. Data from Ref.\,\cite{NDS148} unless indicated otherwise.}
\begin{ruledtabular}
\begin{tabular}{ccccccccccc}

$E_{x,exp.}$ & $E_{x,IBM}$ & $J^{\pi}_i$ & $J^{\pi}_{f,E1}$ & $J_{f,E2}^{\pi}$ & \multicolumn{2}{c}{$B(E1)/B(E2) \left[ 10^{-6} \ \mathrm{fm^{-2}} \right]$} & \multicolumn{2}{c}{$B(E1) \ \left[ 10^{-3} \ \mathrm{e^2fm^2} \right]$} & \multicolumn{2}{c}{$B(E2) \ \left[ \mathrm{W.u.} \right]$} \\
$\left[ \mathrm{keV} \right]$ & $\left[ \mathrm{keV} \right]$ & & & & exp. & IBM & exp.$^{\mathrm{a}}$ & IBM & exp. & IBM \\
\hline
\multicolumn{11}{c}{Members of proposed $0^+_3$ band -- Two-phonon octupole states} \\
1514 & 1601 & $0^+$ & $1^-_1$ & $2^+_1$ & & $8.9 \times 10^4$ & & 17.9 & & $< 0.01$\\
1646 & 1672 & $(2^+)$ & $1^-_1$ & $0^+_1$ & 202(73) & $8.15 \times 10^4$ & & 8.3 & & 0.70 \\
     &      &       & $1^-_1$ & $2^+_1$ & 16(4)  &   $7.35 \times 10^4$ & & & & 0.72\\
1825 & 1757     & $(4^+)$ & $3^-_1$ & $4^+_1$ & 5(2) &  23 & & 5.9 & & 5.6\\
2149 & 1920    & $(6^+)$ & $5^-_1$ & $4^+_1$ &  14(2)$^\mathrm{a}$    & $6.8 \times 10^3$  & $2.49^{+0.18}_{-0.11}$ & 6.6 & 3.7(2)$^\mathrm{a}$ & 0.01 \\
     &      &         & $7^-_1$ & $4^+_1$ &   39(5)$^\mathrm{a}$   &  $1.3 \times 10^4$ & $6.8^{+0.9}_{-1.3}$ & 5.8 & & \\
     &      &         & $5^-_1$ & $6^+_1$ &   0.89(8)$^\mathrm{a}$ &  $1.5 \times 10^3$ & & & 60(7)$^\mathrm{a}$ & 0.04 \\
     &      &         & $7^-_1$ & $6^+_1$ &   2.4(5)$^\mathrm{a}$ &   $2.8 \times 10^3$ & & & & \\
\multicolumn{11}{c}{Members of proposed $0^+_2$ band} \\
917  & 892  & $0^+$ &         & $2^+_1$ &        &        & &  & 31(2) & 39 \\
1171 & 1150 & $2^+$ & $1^-_1$ & $0^+_1$ &        &  10  & & 0.4 & 0.54(7) & 0.8\\
     &      &       & $1^-_1$ & $2^+_1$ &        &  0.3 & & & 14.4(11) & 26\\
1604 & 1789 & $4^+$ & $3^-_1$ & $2^+_1$ & 39(3)$^{\mathrm{a}}$  & $2.3 \times 10^6$ & $0.230^{+0.007}_{-0.009}$ & 1.8 & $0.1252^{+0.0088}_{-0.0006}$$^{\mathrm{a}}$ & $< 0.01$\\
     &      &       & $3^-_1$ & $4^+_1$ & 1.81(13)$^{\mathrm{a}}$ & 3.7 & & & $2.7^{+0.8}_{-0.2}$$^{\mathrm{a}}$& 10\\
2099 & 2507 & $6^+$ & $5^-_1$ & $4^+_1$ & & 183 & & 0.3 & $0.320^{+0.057}_{-0.010}$$^{\mathrm{a}}$& 0.03\\
     &      &       & $7^-_1$ & $4^+_1$ & & 802 & & 1.1 & & \\
     &      &       & $5^-_1$ & $6^+_1$ & & 0.5 & & & $0.3(2)$& 11\\
     &      &       & $7^-_1$ & $6^+_1$ & & 2 & & & & \\
\end{tabular}
\end{ruledtabular}
\begin{flushleft}
    \footnotemark{{\small Calculated from matrix elements reported in Ref.\,\cite{Ibb97a}}}
\end{flushleft}
\end{table*}
\endgroup

\begin{figure}[t]
    \centering
    \includegraphics[width=1\linewidth]{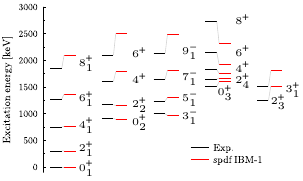}
    \caption{Comparison of bands proposed based on experimental data (black lines) to those obtained with the $spdf$ IBM-1 calculations performed in this work (red lines). Spins were aligned with experimental data. Gray lines connect data to calculated states. Data from Ref.\,\cite{NDS148} unless noted otherwise in the text.}
    \label{fig:comparison_bands}
\end{figure}

To gain further insight, we adopted the $spdf$ IBM-1 parameters reported in Ref.\,\cite{Spi15a, Spi16a} and tried to identify states that could correspond to the experimentally observed states. Calculations were performed with the computer program \textsc{octupole} \cite{Kus}. The comparisons for excitation energies, $B(E1)/B(E2)$ ratios, and $B(E1)$ and $B(E2)$ strengths were added to Table\,\ref{tab:01}. We note that the choice of the Hamiltonian used in Ref.\,\cite{Spi15a, Spi16a} would not lead to $E2$ transitions between states with two negative-parity bosons $(N_{pf} = 2)$ and no negative-parity bosons $(N_{pf} = 0)$ in their wavefunctions. Thus, as in Refs.\,\cite{Lev13a, Spi13a}, we added the dipole interaction with a small coupling strength of $\alpha = 0.001$\,MeV to the Hamiltonian. The value of the latter indicates that $\alpha$ has only a modest impact on general properties of the spectrum, but admixes a small contribution of negative-parity bosons to every state. The admixture of $f$ and $p$ bosons to the ground state of \nuc{148}{Nd} is small, i.e., $N_{pf} = \langle \hat{n}_f \rangle + \langle \hat{n}_p \rangle \approx 0.004$. For the low-spin states, the yrast energies remain almost unchanged. The energies of the $0^+_2$ and $2^+_2$ states change from 888\,keV to 892\,keV and from 1144\,keV to 1150\,keV, respectively. The $3^-_1$ excitation energy decreases from 999\,keV to 968\,keV, and the $1^-_1$ energy from 1023\,keV to 952\,keV. Appreciable mixing happens, however, for the $4^+_2$ and $4^+_3$ states, as well as for the $6^+_2$ and $6^+_3$ states. This mixing pushes the $4^+$ state belonging to the $0^+_2$ band above the two-phonon octupole $4^+$ state, which belongs to the band built on the $0^+_3$ state in these IBM calculations. To ease the further discussion, we summarize the suggested band assignments and the corresponding comparison to the $spdf$ IBM-1 calculations in Fig.\,\ref{fig:comparison_bands}. In the IBM, bands were identified based on enhanced intraband $E2$ transitions between band members with spin $J$ and $J-2$.

Enhanced $B(E1)/B(E2)$ ratios are also obtained in the $spdf$ IBM. As can be seen in Table\,\ref{tab:01} for the comparison of the 1920-keV $6^+$ IBM state and 2149-keV $6^+$ state, the model does a decent job in reproducing the enhanced $E1$ decays. The smaller experimental $B(E1)/B(E2)$ values indicate, however, that the interband $E2$ transitions to the ground-state band are stronger than expected by the current IBM calculations. For instance, the model fails to capture the collectivity of the $6^+ \rightarrow 6^+_1$ transition, see Table\,\ref{tab:01}. Stronger $E2$ transitions could be obtained with more mixing, i.e., a larger value for $\alpha$. This would require significant refitting, which is beyond the scope of this work. The model shows that mixing between the $4^+_2$ $(N_{pf} = 1.4)$ and $4^+_3$ $(N_{pf} =0.6)$ states also leads to enhanced $E1$ transitions for the $4^+$ state belonging to the $0^+_2$ band. The simplistic boson model overestimates this strength, however, by an order of magnitude. $B(E2)$ transition strengths for the $0^+$, $2^+$, and $4^+$ members of the $0^+_2$ band are well reproduced by the model. In contrast to experiment, the model expects some level of collectivity for the $6^+ \rightarrow 6^+_1$ transition of the 2507-keV state rather than for the 1920-keV state.

\begin{figure}[t]
    \centering
    \includegraphics[width=1\linewidth]{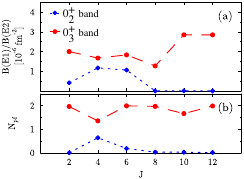}
    \caption{(a) $B(E1;J \rightarrow J-1)/B(E2;J \rightarrow J-2)$ ratios calculated with the $spdf$ IBM-1 for members of the $0^+_2$ (diamonds) and $0^+_3$ (circles) band, respectively. $E1$ transitions lead to members of the one-phonon octupole band, $E2$ transitions correspond to intraband transitions. (b) Number of negative-parity bosons, $N_{pf}$, in the wavefunction of the respective state.}
    \label{fig:be1be2_0+}
\end{figure}

In Fig.\,\ref{fig:be1be2_0+}, we compare theoretical $B(E1;J \rightarrow J-1)/B(E2; J \rightarrow J-2)$ ratios for members of the $0^+_2$ and $0^+_3$ band considering now intraband transitions as discussed in Refs.\,\cite{Bvu13a, Zim16a}. The $E1$ transitions still lead to members of the one-phonon octupole band. As can be seen, the ratios are generally larger for the $0^+_3$ band. With two exceptions, its members have $N_{pf} \approx 2.0$ with the $f$-boson being larger than the $p$-boson component. As already mentioned above, mixing between the $J = 2$, 4, and 6 members of the bands through the added dipole interaction also leads to enhanced $B(E1)/B(E2)$ ratios for the $0^+_2$ band members. Beyond $J=8$, mixing between the two bands is negligible in the model and the $B(E1)$ strength for members of the $0^+_2$ band drops significantly. The $10^+$ state of the $0^+_3$ band has $N_{pf} < 2.0$ since it is mixing with $10^+$ member of the ground-state band.

For completeness, and before discussing the $(p,t)$ intensities, we comment on some $B(E3)$ reduced transition strengths as these were used in Ref.\,\cite{Ibb97a} to argue in favor of a two-phonon octupole structure. First, we consider only the $[ s^{\dagger} \Tilde{f} + f^{\dagger} s ]^{(3)}$ part of the $E3$ operator defined in Ref.\,\cite{Zam03a}, setting all other terms to zero. We set the effective charge $e_{3,sf} = 0.115 \ \mathrm{eb^3}$ to match the $B(E3;3^-_1 \rightarrow 0^+_1)$ strength of 34(3) W.u. \cite{Kib02a}. With that choice, the IBM predicts $B(E3;6^+_{1920} \rightarrow 3^-_1) = 77$ W.u., $B(E3;6^+_{1920} \rightarrow 5^-_1) = 12$ W.u., and $B(E3;6^+_{1920} \rightarrow 7^-_1) = 2$ W.u. for the $E3$ transitions originating from the two-phonon octupole $6^+$ state. Within their comparably large uncertainties, the experimental values $B(E3;6^+_{2149} \rightarrow 3^-_1) = 17^{+43}_{-17}$ W.u., $B(E3;6^+_{2149} \rightarrow 5^-_1) = 34^{+92}_{-28}$ W.u., and $B(E3;6^+_{2149} \rightarrow 7^-_1) = 27^{+120}_{-17}$ W.u. are consistent with the model expectations for the corresponding transitions \cite{Ibb97a}. However, the last two values could be significantly larger possibly indicating that other terms of the $E3$ operator would be needed to reproduce them. An inspection of the associated matrix elements suggests that this could be achieved by including the $[ d^{\dagger} \Tilde{f} + f^{\dagger} \Tilde{d} ]^{(3)}$ term, which is larger for the transitions coming from the two-phonon $6^+$ state than for the $3^-_1 \rightarrow 0^+_1$ transition. Choosing $e_{3,sf} = 0.099  \ \mathrm{eb^3}$ and $e_{3,df} = -0.12  \ \mathrm{eb^3}$ maintains the agreement for the $B(E3;3^-_1 \rightarrow 0^+_1)$ strength and yields $B(E3;6^+_{1920} \rightarrow 3^-_1) = 88$ W.u., $B(E3;6^+_{1920} \rightarrow 5^-_1) = 29$ W.u., and $B(E3;6^+_{1920} \rightarrow 7^-_1) = 7$ W.u., possibly in better overall agreement with experiment. For the $8^+$, which was tentatively placed at 2726\,keV, Ibbotson {\it et al.} determined $B(E3;8^+_{2726} \rightarrow 5^-_1) = 65(64)$\,W.u.\,\cite{Ibb97a}. As indicated in the discussion around Fig.\,\ref{fig:be1be2_0+}, significantly less mixing is expected for the two-phonon octupole $8^+$ state. With the second set of parameters, the IBM yields $B(E3;8^+_{2324} \rightarrow 5^-_1) = 77$\,W.u. for the $E3$ transition from the two-phonon octupole $8^+$ state to the one-phonon octupole $5^-$ state. This value also agrees with experiment within the large uncertainties reported in Ref.\,\cite{Ibb97a}. We note that the 2726-keV $8^+$ state would fit into the band built on the $0^+_3$ state, see Fig.\,\ref{fig:rotation}. 

Ibbotson {\it et al.} also determined $E3$ matrix elements for the 1604-keV state \cite{Ibb97a}, i.e., for the $4^+$ state where we claim that it does not belong to the two-phonon octupole band. To support this claim, we compare the $B(E3)$ strengths for the $4^+ \rightarrow 1^-_1$ and $4^+ \rightarrow 3^-_1$ transitions to the IBM predictions. Strengths of $B(E3;4^+_{1604} \rightarrow 1^-_1) = 6.7^{+17.3}_{-6.7}$\,W.u. and $B(E3;4^+_{1604} \rightarrow 3^-_1) \leq 12.3$\,W.u. are determined from the $E3$ matrix elements reported in Ref.\,\cite{Ibb97a}. With the second set of parameters, the IBM predicts 13.5 and 10.6\,W.u., respectively. These values are consistent with the experimental data. As mentioned above, the reason for the $E3$ transitions to be observed for the $sd$-dominated $4^+$ state is the mixing between the $4^+_2$ and $4^+_3$ states, see also discussion around Fig.\,\ref{fig:be1be2_0+}, induced by the dipole interaction. One might, thus, expect that the $E3$ matrix elements for the 1825-keV state are even larger.

\begin{figure}
    \centering
    \includegraphics[width=1\linewidth]{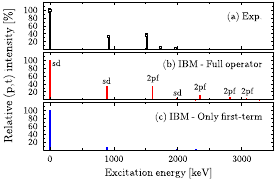}
    \caption{(a) Experimental and (b) and (c) $spdf$ IBM-1 distribution of the relative $(p,t)$ transfer strength for (firmly assigned) $0^+$ states in \nuc{148}{Nd}. (b) Strengths obtained with full operator of Eq. (\ref{eq:01}). (c) Strengths obtained when only the first part of the operator of Eq. (\ref{eq:01}) is used. Strengths were normalized to the ground-state to ground-state transfer strength. In panel (b), $sd$ indicates a dominant quadrupole structure of the $0^+$ state, while $2pf$ means that two negative-parity bosons are part of the wavefunction of the $0^+$ state.}
    \label{fig:intensities}
\end{figure}

We now turn to the relative $(p,t)$ intensities and their description with the $spdf$ IBM-1. The full comparison is shown in Fig.\,\ref{fig:intensities}. For \nuc{150}{Nd}, the parameters reported in Refs. \cite{Spi15a, Spi16a} were also used and the dipole term with the same strength as for \nuc{148}{Nd} added to the Hamiltonian. We adopted a similar two-neutron transfer operator as in Refs.\,\cite{Lev13a, Spi13a}, but chose to include the $\alpha_d \hat{n}_d \hat{s}$ term as well:

\begin{align}
    \hat{P}^{(0)}_{-,\nu} = &\alpha_{\nu} \left( \Omega_{\nu} - N_{\nu} - \frac{N_{\nu}}{N} \hat{n}_d \right)^{1/2} \left( \frac{N_{\nu}+1}{N+1} \right)^{1/2} \hat{s} \nonumber \\
    &+ \left( \alpha_p \hat{n}_p + \alpha_f \hat{n}_f + \alpha_d \hat{n}_d \right) \hat{s}
    \label{eq:01}
\end{align}

To obtain the relative $(p,t)$ transfer intensities shown in Fig. \ref{fig:intensities}\,(b), the following parameters were used: $\Omega_{\nu} = (82-50)/2=16$, $N_{\nu} = (88-82)/2=3$, $\alpha_p = \alpha_f = 2.5$, $\alpha_d = 0.08$, and $\alpha_{\nu} = 0.11$. Fig.\,\ref{fig:intensities}\,(c) shows the relative $(p,t)$ intensities obtained with the $spdf$ IBM-1 when $\alpha_p = \alpha_f = \alpha_d = 0$. In this work, we use dimensionless parameters as only intensities relative to the ground-state to ground-state transfer strength are given. The agreement between the experimental and theoretical strength patterns when including the additional terms in the operator is excellent. For the present IBM calculations, including these terms is clearly necessary as can be seen by comparing Figs.\,\ref{fig:intensities}\,(b) and (c). While the $0^+_2$ state obtains its strength through the $\alpha_d \hat{n}_d \hat{s}$ term (only 6\,\% without it), the strength of the $0^+_1(\nuc{150}{Nd}) \rightarrow 0^+_3(\nuc{148}{Nd})$ transfer results entirely from the $(\alpha_p \hat{n}_p + \alpha_f \hat{n}_f)\hat{s}$ term of the operator in Eq.\,(\ref{eq:01}). We note that setting $\alpha_p = 0$ and increasing $\alpha_f$ by about a factor of three would lead to similar results. This would correspond to a similar parameter choice as presented in Ref.\,\cite{Nom26a}. Obviously, $\alpha_p$ and $\alpha_f$ are significantly larger than the other parameters that act predominantly on $sd$ states. This was also observed in Refs.\,\cite{Lev13a, Spi13a}. The two parameters, $\alpha_p$ and $\alpha_f$, could be smaller if the strength of the dipole interaction in the Hamiltonian was larger, which would admix more negative-parity bosons to the wavefunctions of dominant $sd$ states. As mentioned earlier, larger mixing could also improve the agreement for some of the reduced transition strengths presented in Table\,\ref{tab:01}. It would require a refitting of both nuclei though beyond the already extensive studies of rare-earth nuclei and associated structure observables in Refs.\,\cite{Spi15a, Spi16a}. In any case, in combination with the discussion around Table\,\ref{tab:01}, the agreement obtained between experiment and theory for the relative $(p,t)$ intensities also supports a significant contribution of the two-phonon octupole structure to the wavefunction of the $0^+_3$ state and for members of its associated band. The schematic calculations prove that under certain conditions $(p,t)$ reactions can strongly populate such states. We note that the wavefunction of the $0^+_3$ state also has a significant $p$-boson component, i.e., $\langle \hat{n}_p \rangle  =0.7$ and $\langle \hat{n}_f \rangle  =1.3$. For completeness, we add that $\langle \hat{n}_p \rangle  =0.18$ and $\langle \hat{n}_f \rangle  =0.82$ for the $3^-_1$ state, and $\langle \hat{n}_p \rangle  =0.47$ and $\langle \hat{n}_f \rangle  =0.53$ for the $1^-_1$ state. Both one-phonon octupole states, therefore, have already a sizable $p$-boson contribution. A possible interpretation of the $p$ boson was provided in Ref.\,\cite{Spi15a} and references therein.

\begin{figure}
    \centering
    \includegraphics[width=1\linewidth]{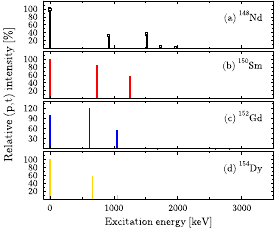}
    \caption{Relative $(p,t)$ intensities for the stable $N=88$ isotones. Data are taken from Ref.\,\cite{Deb72a} for \nuc{150}{Sm}, Ref.\,\cite{Mey06a} for \nuc{152}{Gd}, and Ref.\,\cite{Qu92a} for \nuc{154}{Dy}. Note that for the other $N=88$ isotones, the relative $(p,t)$ intensities were determined at one specific scattering angle rather than through the angle-integrated cross sections as done in this work.}
    \label{fig:intensities_88}
\end{figure}

We compare the available $(p,t)$ data for the $N=88$ isotones \cite{Deb72a, Mey06a, Qu92a} in Fig.\,\ref{fig:intensities_88}. As can be seen, the relative strength distribution is different for all isotones. However, with the exception of \nuc{154}{Dy}, the $0^+_3$ state shows appreciable $(p,t)$ transfer strength across the $N=88$ isotones. It appears to peak for \nuc{150}{Sm} and \nuc{152}{Gd} with a relative strength of around 55\,\% of the ground-state to ground-state transfer strength. Note that in \nuc{152}{Gd} and \nuc{154}{Dy}, the $0^+_3$ state is lower in energy than the $1^-_1$ and $3^-_1$ states. It is, thus, unlikely that this state corresponds to a two-phonon octupole vibrational state in these nuclei. In \nuc{150}{Sm}, the $0^+_3$ state is energetically very close to the one-phonon octupole vibrational states. Pronounced anharmonicities would need to be present to explain a significant two-phonon octupole admixture to this $0^+$ state's wavefunction. For completeness, we note that Ref.\,\cite{Nom26a} predicted that all $0^+_2$ states of the $N=88$ isotones should be two-phonon octupole states. The excitation energies of the $0^+_2$ states were systematically overpredicted though, sometimes with energy differences larger than 1\,MeV, when determining the $sdf$ IBM-1 parameters by mapping the IBM potential energy surface to the one predicted by Gogny-HFB calculations \cite{Nom26a}. Likely, the two-phonon octupole $0^+$ states are at higher energies in the other $N=88$ isotones. If one observes the same level of anharmonicity as in \nuc{148}{Nd}, then the first two-phonon octupole $0^+$ states would be expected at approximately 1623\,keV, 1702\,keV, and 1830\,keV for \nuc{150}{Sm}, \nuc{152}{Gd}, and \nuc{154}{Dy}, respectively. For \nuc{154}{Dy}, no $0^+$ states with energies higher than the $0^+_3$ state have been reported yet. For \nuc{150}{Sm}, Schmelzenbach {\it et al.} identified a $0^+$ state at 1603\,keV based on $\gamma \gamma$ angular correlation studies following the $\beta$ decay of \nuc{150}{Pm} and \nuc{150}{Eu$^m$} \cite{Sch18a}. This state, which is remarkably close to the estimated energy of a possible two-phonon octupole $0^+$ state, has a significantly enhanced $B(E1;0^+ \rightarrow 1^-_1)/B(E2;0^+ \rightarrow 2^+_1)$ ratio of $69(4) \times 10^{-6} \ \mathrm{fm^{-2}}$. For \nuc{152}{Gd}, a $0^+$ state is adopted at 1681\,keV which also decays via an $E1$ transition to the $1^-_1$ state \cite{Mar13a}. It is remarkably close to the estimated energy as well. As for \nuc{150}{Sm}, the observed $B(E1;0^+ \rightarrow 1^-_1)/B(E2;0^+ \rightarrow 2^+_1)$ ratio is enhanced with $16.0(11) \times 10^{-6}$\,fm$^{-2}$ \cite{Mar13a}. Its relative $(p,t)$ strength is smaller than 1\,\% though \cite{Mey06a}, clearly pointing out that two-phonon octupole $0^+$ states are not always strongly populated in $(p,t)$ reactions and that their $(p,t)$ strength likely depends on where they appear in the spectrum. More experimental data are certainly needed to uniquely identify the two-phonon octupole states at and beyond $N=88$, and to better understand their population in $(p,t)$ reactions.

\begin{figure}
    \centering
    \includegraphics[width=1\linewidth]{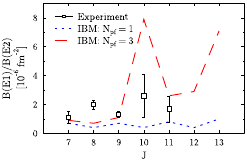}
    \caption{$B(E1;J \rightarrow J-1)/B(E2;J \rightarrow J-2)$ ratios for members of the ground-state and first negative-parity band in \nuc{148}{Nd} compiled from the adopted data \cite{NDS148}. A comparison to $spdf$ IBM-1 calculations, mentioned in the text, is shown when only allowing for one negative-parity boson ($N_{pf} =1$) and when allowing for up to three negative-parity bosons ($N_{pf} =3$) in the wavefunctions.}
    \label{fig:be1be2}
\end{figure}

In closing, we want to acknowledge that Bvumbi {\it et al.} earlier argued for a shape-coexistence scenario, where the $0^+_2$ band and negative-parity band of \nuc{150}{Sm} and \nuc{152}{Gd} form an alternating parity band \cite{Bvu13a}. In that case, the $0^+_2$ band belongs to a configuration that is octupole deformed and different from the ground-state configuration. They also provided some evidence, based on the analysis of the rotational frequencies of the $0^+_2$ band and negative-parity band, that such an octupole-deformed band might stabilize around $J = 6$, highlighting at the same time the marked change in slope of their calculated Routhians between $J=4$ and $J=6$ for the $0^+_2$ band of \nuc{150}{Sm}. These are the spins for which we discussed pronounced mixing between the $sd$-dominated and $2pf$-dominated states in \nuc{148}{Nd}. We do, however, acknowledge that the parabolic evolution of the $0^+_2$ and $0^+_3$ states' excitation energies with a minimum at $Z=64$, which can be seen in Fig.\,\ref{fig:intensities_88}, could indeed be an indicator of shape coexistence in the $N=88$ isotones. A scenario favoring the condensation of rotational-aligned octupole phonons \cite{Fra08a} instead of shape coexistence with an octupole-deformed configuration, which Bvumbi {\it et al.} also mentioned but ruled out for \nuc{152}{Gd}, was presented in Ref.\,\cite{Spi15b} for \nuc{146}{Nd}. In that work, bands possibly corresponding to zero-, one-, two-, and three-phonon octupole excitations were identified and a case was made that the two- and three-phonon octupole configurations are becoming yrast at around $J^{\pi} = 10^+$ \cite{Spi15b}. It was shown that the $spdf$ IBM-1 was able to describe this structure change in \nuc{146}{Nd} through the study of the signature splitting between the respective positive- and negative-parity bands without invoking shape coexistence. Such data are not available for \nuc{148}{Nd} to perform a similar study. The available data allow us, however, to study the $B(E1)/B(E2)$ ratios for members of the ground-state and negative-parity band with $7 \leq J \leq 11$. The comparison to the $spdf$ IBM-1 calculations including only one negative-parity boson $(N_{pf} =1)$ and allowing for up to three negative-parity bosons $(N_{pf}=3)$ as part of the wavefunctions is shown in Fig.\,\ref{fig:be1be2}. An even-odd staggering is observed, where the ratios are larger for the even-$J$ states. In conflict with the data, the IBM expects that the ratios are larger for the odd-$J$ states when only one negative-parity boson is included in the calculations. When allowing for up to three negative-parity bosons, the IBM gets the $B(E1)/B(E2)$ ratios for the odd-$J$ states spot on. It fails to describe the two known experimental ratios for positive-parity states though, underestimating the ratio for the $8^+$ state and overestimating the ratio for the $10^+$ state. Nevertheless, when only inspecting the $9 \leq J \leq 11$ sequence, the correct even-odd staggering is reproduced. The appearance of this staggering can be traced back to a change in the states' wavefunctions with the two-phonon octupole structure becoming yrast at $J^{\pi} = 10^+$ as it did in the IBM calculations for \nuc{146}{Nd} \cite{Spi15b}. Another very intriguing feature is observed in Fig.\,\ref{fig:be1be2} at $J=13$, where the three-phonon octupole states become yrast for the negative-parity states and the staggering order changes again. Future experiments could try to look for this change in staggering of the $B(E1)/B(E2)$ ratios in the yrast sequence to possibly test the appearance of bands with multiple octupole phonons further or to search for signatures of a configuration that is indeed octupole deformed.

\section{Summary and Outlook}

We performed a two-neutron $(p,t)$ transfer experiment at the FSU John D. Fox Accelerator Laboratory with the SE-SPS to study excited states of \nuc{148}{Nd} close to the $N=90$ phase-transitional point and $N=88$ where enhanced octupole correlations are expected. A total of 54 excited states of \nuc{148}{Nd} were identified up to an excitation of about 3500\,keV and $(p,t)$ angular distributions measured.

In this publication, we focused on $J^{\pi} = 0^+$ states and their $J^{\pi} = 2^+$ and $4^+$ band members, which were populated in the $(p,t)$ reaction. We provided an extensive review of their previously reported $\gamma$-decay properties and compared these to $spdf$ IBM-1 calculations. The detailed comparison supports that the $0^+_3$ state and its band members can be interpreted as two-phonon octupole states. Based on this interpretation, we also propose to assign the 2149-keV $6^+$ state to the band built on the $0^+_3$ state rather than to the one built on the $0^+_2$ state. Our work establishes that contrary to previous claims \cite{Ibb97a, Nom21a, Nom26a} it is not the $0^+_2$ state and its band members which show enhanced octupole correlations in \nuc{148}{Nd}. The IBM calculations, presented in this work, also provide a satisfactory description of the previously observed enhanced $E3$ transitions of the two-phonon octupole $6^+$ and $8^+$ candidates to the negative-parity band \cite{Ibb97a}.

We compared our new $(p,t)$ data for $0^+$ states in \nuc{148}{Nd} to data available for the other stable $N=88$ isotones. Besides a parabolic evolution of the $0^+_2$ and $0^+_3$ states' excitation energies with a minimum at $Z=64$, which could be a hint for shape coexistence, we pointed out that the $0^+_2$ and $0^+_3$ states of the other $N=88$ isotones likely are not two-phonon octupole states. Instead, we proposed possible candidates in \nuc{150}{Sm} and \nuc{152}{Gd} based on a comparison to \nuc{148}{Nd} and already observed enhanced $B(E1)/B(E2)$ ratios for these candidates. We also showed that possible mixing between $sd$-dominated states and states with $N_{pf} =2$ could lead to enhanced $B(E1)/B(E2)$ ratios for the $sd$-dominated states.

In the last part of this work, the $B(E1)/B(E2)$ ratios in the yrast sequence of \nuc{148}{Nd} were compared to IBM predictions. We found that the even-odd staggering and changes of it could be indicators of admixtures of configurations with multiple octupole phonons to the yrast states. We proposed that a measurement of these ratios for states with $J \geq 11$ could likely test whether as in \nuc{146}{Nd} \cite{Spi15b} the two-phonon and three-phonon octupole states become yrast at moderate spin. A larger experimental database will likely also be able to test whether one observes bands with different numbers of octupole phonons or whether there is indeed a configuration with enhanced octupole correlations that is different from the ground-state configuration, and that could possibly be connected to the findings of Ref.\,\cite{Gar09a} for \nuc{152}{Sm}. The additonal negative-parity band supposedly built on the configuration belonging to the $0^+_2$ state \cite{Gar09a} was successfully reproduced in Ref.\,\cite{Xia26a}. In their work using a newly developed triaxial quadrupole-octupole collective Hamiltonian \cite{Xia26a}, Xiang {\it et al.} stated that the $0^+_2$ state corresponded to a classical $\beta$ vibration, therefore indirectly implying that a shape-coexistence scenario does not seem necessary. Further systematic studies in this mass region might be necessary to arrive at a definite conclusion.

\begin{acknowledgments}
This work was supported by the U.S. National Science Foundation under Grants No. PHY-2012522 (FSU) and No. PHY-2412808 (FSU), as well as by the Department of Energy, National Nuclear Security Administration, under Award No. DE-NA0004150 through the Center for Excellence in Nuclear Training And University-based Research (CENTAUR). Additional support by Florida State University is gratefully acknowledged. This research used targets provided by the Center for Accelerator Target Science at Argonne National Laboratory, a DOE Office of Science User Facility supported by the U.S. Department of Energy, Office of Nuclear Physics, under Award No. DE-AC02-06CH11357. A.L.C. and M.S. thank K.W. Kemper for many inspiring discussions, P.D. Cottle for comments on the manuscript, and S. Pascu for providing the $spdf$ IBM-1 computer program.

\end{acknowledgments}

\bibliography{references}

\end{document}